\documentclass[prd,superscriptaddress,nofootinbib,amsmath,amssymb,aps,11pt]{revtex4-2}

\usepackage{bm}
\usepackage{amsfonts}
\usepackage{latexsym}
\usepackage[latin1]{inputenc}
\usepackage{graphicx}
\usepackage{amsmath}
\usepackage{palatino}
\usepackage{mathpazo}
\usepackage[normalem]{ulem}
\usepackage{xcolor}

\usepackage{booktabs}
\usepackage{dcolumn}

\usepackage{natbib}
\usepackage{color,soul}

\begin{document}
\title{The power of binary pulsars in testing Gauss-Bonnet gravity}

\author{Petar Y. Yordanov}
\email{pyordanov@phys.uni-sofia.bg}
\affiliation{Department of Theoretical Physics, Faculty of Physics, Sofia University, Sofia 1164, Bulgaria}

\author{Kalin V. Staykov}
\email{kstaykov@phys.uni-sofia.bg}
\affiliation{Department of Theoretical Physics, Faculty of Physics, Sofia University, Sofia 1164, Bulgaria}

\author{Stoytcho S. Yazadjiev}
\email{yazad@phys.uni-sofia.bg}
\affiliation{Theoretical Astrophysics, Eberhard Karls University of T\"ubingen, T\"ubingen 72076, Germany}
\affiliation{Department of Theoretical Physics, Faculty of Physics, Sofia University, Sofia 1164, Bulgaria}
\affiliation{Institute of Mathematics and Informatics, 	Bulgarian Academy of Sciences, 	Acad. G. Bonchev St. 8, Sofia 1113, Bulgaria}

\author{Daniela D. Doneva}
\email{daniela.doneva@uni-tuebingen.de}
\affiliation{Theoretical Astrophysics, Eberhard Karls niversity of T\"ubingen, T\"ubingen 72076, Germany}
\affiliation{INRNE - Bulgarian Academy of Sciences, 1784  Sofia, Bulgaria}

\date{}

\newcommand{\ks}[1]{\textcolor{purple}{[KS: #1]}}

\begin{abstract}
    Binary pulsars are a powerful tool for probing strong gravity that still outperforms direct gravitational wave observations in a number of directions due to the remarkable accuracy of the pulsar timing. They can constrain very precisely the presence of additional charges of the orbiting neutron stars leading to new channels of energy and angular momentum loss, such as the scalar dipole radiation. In the present paper, we explore in detail the possibility of constraining different classes of scalar-Gauss-Bonnet gravity with binary pulsars. Additionally, the existing constraints related to the observed maximum mass of neutron stars are also updated. Interestingly, depending on the equation of state, the resulting limits on the theory coupling parameters can outperform the constraints coming from binary merger observations by up to a factor of 2 even for the so-called Einstein-dilaton-Gauss-Bonnet gravity where neutron stars are often underestimated as relevant theory probes. As an additional merit, precise Bayesian methods are compared with approximate approaches with the latter showing very good performance despite their simplicity. 
\end{abstract}

\maketitle

\section{Introduction}

Binary pulsars are among the first systems to allow testing general relativity (GR) in the strong field regime \cite{Damour:1990wz,Damour:1991rd,Esposito-Farese:1996kbb,Damour:1996ke,Weisberg:2004hi,Freire:2022wcz}. Different post-Keplerian parameters in the orbital motion of the neutron stars can be measured \cite{Will:2014kxa,Freire:2000ey} and one of the most intriguing is related to the shrinking of the binary orbit due to gravitational wave emission. The observations fit very well the GR predictions and that is why modified theories of gravity possessing an additional channel of energy emission, attributed for example to the existence of a new fundamental field, are severely constrained \cite{Kramer:2021jcw}. A prominent example of a theory, practically ruled out by the binary pulsar observations, is the Damour-Esposito-Farese (DEF) model which is a subclass of the massless scalar-tensor theories (STT) \cite{Damour:1992we, Damour:1993hw,Shao:2017gwu,Chiba:2021rqa,Zhao:2022vig} \footnote{Note that a nonzero scalar field mass can evade the binary pulsar constraints \cite{Ramazanoglu:2016kul,Yazadjiev:2016pcb,Rosca-Mead:2020bzt}, rapid rotation can magnify the deviations from GR considerably \cite{Doneva:2013qva}, and other sectors of STTs are only weakly constrained by binary pulsars \cite{Mendes:2016fby,Mendes:2019zpw}.}. It outperforms by far the constraints based directly on the neutron star mass and radius observations \cite{Tuna:2022qqr}. In the DEF models, the additional channel of energy loss is controlled by the scalar charge (the coefficient in front of the leading order $1/r$ scalar field asymptotic at infinity) leading to scalar dipole radiation.  Constraints on the DEF model were confirmed by a Bayesian analysis that has employed a more sophisticated equation of state and post-Keplerian parameters treatment \cite{Anderson:2019eay} as well as new pulsar timing models \cite{Batrakov:2023imd}. Interestingly, a careful comparison in \cite{Anderson:2019eay} revealed that the sophisticated Bayesian approach leads to very similar results as the simpler treatment employed in the past \cite{Damour:1996ke,Freire:2012mg}.

Another well-motivated class of extended STTs in which neutron star solutions were well-studied is the scalar-Gauss-Bonnet (sGB) gravity. Being more complicated than the DEF model due to the addition of a second-order (Gauss-Bonnet) curative invariant, it can be considered as an effective field theory being motivated by quantum gravity. It brings new interesting phenomenology such as the violation of the black hole no-scalar hair theorems \cite{Mignemi:1992nt,Kanti:1995vq,Torii:1996yi,Pani:2009wy}. Two major classes of sGB theories became popular in the literature. The first one is the shift symmetric theory with a linear coupling between the scalar field and the Gauss-Bonnet invariant for which black holes are always endowed by a scalar field. It can be considered as a leading order expansion of the so-called Eintein-dilaton-Gauss-Bonnet (EdGB) gravity in the case of a weak coupling  \cite{Mignemi:1992nt,Kanti:1995vq,Torii:1996yi,Pani:2009wy}. Neutron star solutions in these theories were constructed both in the static and rapidly rotating regimes \cite{Pani:2011xm,Kleihaus:2014lba,Kleihaus:2016dui}. Another interesting class of sGB theories admits the so-called curvature induces spontaneous scalarization  \cite{Doneva:2017bvd,Silva:2017uqg,Antoniou:2017acq}, analogous to the neutron star spontaneous scalarization in the DEF model \cite{Damour:1992we}. Neutron stars in this case were obtained only in the static limit \cite{Silva:2017uqg,Doneva:2017duq,Xu:2021kfh}.

Gravitational wave constraints on sGB gravity were addressed mainly in the context of  black hole-black hole \cite{Perkins:2021mhb,Wong:2022wni,Wang:2021jfc,Tahura:2018zuq,Yamada:2019zrb} and black hole-neutron star \cite{Lyu:2022gdr} merger events. Constraints of sGB gravity through binary pulsars are still scarcely studied with the only exception being \cite{Danchev:2021tew} where the sGB gravity admitting spontaneous scalarization was adopted and \cite{Yagi:2015oca} where a EdGB theory with a particular coupling function was considered. One of the reasons is that in the shift-symmetric sGB gravity, neutron stars do not possess a scalar charge \cite{Yagi:2015oca} thus the orbital decay will be the same as in GR. Moving from a linear scalar field coupling, like in the shift-symmetric case, to the exponential coupling in the EdGB gravity eventually leads to a scalar charge development (see e.g. \cite{Kleihaus:2014lba,Kleihaus:2016dui}) that alters the binary dynamics. Studying the induced constraints on EdGB gravity is one of the goals of the present paper. 

Contrary to previous studies \cite{Danchev:2021tew} we will not perform a full Bayesian analysis, but instead a simpler and more tractable algorithm with be developed for deriving constraints similar to the original studies in the DEF model \cite{Damour:1996ke,Freire:2012mg}. Estimating the power of this treatment is another goal.   

A complementary approach for constraining the sGB gravity is through the existence of solutions. It is well known that neutron stars in sGB gravity possess a lower maximum mass compared to GR (for the same equation of state) \cite{Pani:2011xm,Doneva:2017duq}. This allows us to set additional constraints on the parameters of the theory by taking into account the maximal observed neutron star mass up to now \cite{Pani:2011xm}.  We will update previous results by considering the most recent neutron star observations with higher neutron star masses and compare them to the constraints coming from binary pulsars.

The paper is structured as follows: In Section II we shortly introduce the Gauss-Bonnet gravity. We continue in Section III with the description of the two methods we adopted for constraining the theory. The results are presented in Section IV. The paper ends with a Conclusion. 

\section{scalar-Gauss-Bonnet gravity}
In the present paper, we study neutron stars in scalar-Gauss-Bonnet gravity. The general form of the action can be written as:  
\begin{eqnarray}
S=&&\frac{1}{16\pi}\int d^4x \sqrt{-g} 
\Big[R - 2\nabla_\mu \varphi \nabla^\mu \varphi  - V(\varphi)
+ \lambda^2 f(\varphi){\cal R}^2_{GB} \Big] + S_{\rm matter}  (g_{\mu\nu},\chi) ,\label{eq:quadratic}
\end{eqnarray}
where $R$ is the Ricci scalar and $\nabla_{\mu}$ is the covariant derivative with respect to the metric $g_{\mu\nu}$. $V(\varphi)$ is the scalar field potential and  $f(\varphi)$ controls the dimensionless part of the coupling between the scalar field  $\varphi$ and the Gauss-Bonnet invariant ${\cal R}^2_{GB}=R^2 - 4 R_{\mu\nu} R^{\mu\nu} + R_{\mu\nu\alpha\beta}R^{\mu\nu\alpha\beta}$. Throughout this paper, we will assume $V(\varphi) = 0$ while the particular choices of $f(\varphi)$ will be commented in detail below. The Gauss-Bonnet coupling constant $\lambda$ has dimension of $length$. $S_{\rm matter}$ is the matter action. 

The field equations derived from the action (\ref{eq:quadratic}) are 
\begin{eqnarray}\label{FE}
&&R_{\mu\nu}- \frac{1}{2}R g_{\mu\nu} + \Gamma_{\mu\nu}= 2\nabla_\mu\varphi\nabla_\nu\varphi -  g_{\mu\nu} \nabla_\alpha\varphi \nabla^\alpha\varphi -\frac{1}{2}g_{\mu\nu}V(\varphi)  + 8\pi T^{\rm matter}_{\mu\nu},\\
\label{eq:fe_scalar}
&&\nabla_\alpha\nabla^\alpha\varphi= \frac{1}{4}\frac{dV(\varphi)}{d\varphi}  -  \frac{\lambda^2}{4} \frac{df(\varphi)}{d\varphi} {\cal R}^2_{GB},
\end{eqnarray}
where  $T^{\rm matter}_{\mu\nu}$ is the matter energy momentum tensor.  $\Gamma_{\mu\nu}$ is defined by 
\begin{eqnarray}
\Gamma_{\mu\nu}&=& - R(\nabla_\mu\Psi_{\nu} + \nabla_\nu\Psi_{\mu} ) - 4\nabla^\alpha\Psi_{\alpha}\left(R_{\mu\nu} - \frac{1}{2}R g_{\mu\nu}\right) + 
4R_{\mu\alpha}\nabla^\alpha\Psi_{\nu} + 4R_{\nu\alpha}\nabla^\alpha\Psi_{\mu} \nonumber \\ 
&& - 4 g_{\mu\nu} R^{\alpha\beta}\nabla_\alpha\Psi_{\beta} 
+ \,  4 R^{\beta}_{\;\mu\alpha\nu}\nabla^\alpha\Psi_{\beta} 
\end{eqnarray}  
with 
\begin{eqnarray}
\Psi_{\mu}= \lambda^2 \frac{df(\varphi)}{d\varphi}\nabla_\mu\varphi .
\label{eq:Psi_mu}
\end{eqnarray}

The equation for hydrostatic equilibrium of the fluid which can be derived  from the  Bianchi identity
\begin{eqnarray}\label{BID}
\nabla^{\mu}T^{\rm matter}_{\mu\nu}=0. 
\end{eqnarray}

In the present paper, we will study static and spherically symmetric spacetime and static and spherically symmetric scalar field and fluid configuration. For the spacetime metric we adopt the standard ansatz:
\begin{eqnarray}
\label{eq:metric}
ds^2= - e^{2\Phi(r)}dt^2 + e^{2\Lambda(r)} dr^2 + r^2 (d\theta^2 + \sin^2\theta d\phi^2 ).
\end{eqnarray}   

We choose the matter source to be a perfect fluid with $T^{\rm matter}_{\mu\nu}=(\rho + p)u_{\mu}u_{\nu} + pg_{\mu\nu}$ 
where $\rho$, $p$ and $u^{\mu}$ are the energy density, pressure, and 4-velocity of the fluid, respectively.  For the explicit form of the dimensionally reduced field equations and the equation for hydrostatic equilibrium, we refer the interested reader to \cite{Doneva:2017duq}.  

To model a neutron star, an equation of state (EoS) should be specified. In the present paper, we use different realistic EoS represented through their piecewise polytropic approximation \cite{Read:2008iy}. The specific EoSs we employ will be commented on below. 

The boundary conditions are the natural ones -- regularity at the center of the star:

\begin{equation}
	\left.\Lambda\right|_{r\rightarrow 0}\rightarrow 0,\quad \left. \frac{d\Phi}{dr}\right|_{r\rightarrow 0}\rightarrow 0, \quad \left.\frac{d\varphi}{dr}\right|_{r\rightarrow 0}\rightarrow 0,
\end{equation}
and asymptotic flatness at spacial infinity:
\begin{equation}
	\left.\Lambda\right|_{r\rightarrow \infty}\rightarrow 0,\quad \left.\Phi\right|_{r\rightarrow \infty}\rightarrow 0, \quad \left.\varphi\right|_{r\rightarrow \infty}\rightarrow 0.
\end{equation}
Typically in sGB gravity, the regularity at the stellar center fails to be fulfilled for large central energy densities. Therefore, the scalarized branches of solutions are terminated at a fixed maximum value of $\rho$. This is governed by the following regularity condition
\begin{equation}\label{eq:SolutionConstrain}
9  \lambda^4  \left(\frac{df}{d\varphi}(\varphi_0)\right)^2 (\Lambda_2)^4 - 72 \pi \lambda^4 p_0  \left(\frac{df}{d\varphi}(\varphi_0)\right)^2   (\Lambda_2)^3 - 6\pi \rho_0  \Lambda_2 + 16 \pi^2 \rho_{0}^2=0,
\end{equation}
where $\varphi_0, \rho_0, p_0, \Lambda_2$ are the coefficients in expansion at the stellar center
\begin{equation}
\Lambda=\Lambda_0 + \Lambda_1 r+\frac{1}{2} \Lambda_2 r^2  + O(r^3);\;\Phi=\Phi_0+\Phi_1 r+\frac{1}{2} \Phi_2 r^2 + O(r^3);\; \varphi=\varphi_0+\varphi_1 r + \frac{1}{2} \varphi_2 r^2 + O(r^3).
\end{equation}
Eq. \eqref{eq:SolutionConstrain} constitutes a fourth-order algebraic equation for $\Lambda_2$ that depends on the central values of the pressure $p_0$, the energy density $\rho_0$, and the scalar field $\varphi_0$. 
In case no real roots for $\Lambda_2$ exist, there are no regular neutron stars.

\subsection{The coupling function}
We will be interested in two flavors of sGB theories. The first one is the EdGB gravity having $\frac{df}{d\varphi}(0) \ne 0$. More specifically, we will focus on 
\begin{equation}
    f_{\rm EdGB_1}(\varphi) = \frac{1}{2\beta}e^{2 \beta \varphi},
    \label{eq:coupling_f3}
\end{equation}
where $\beta$ is a parameter. According to eq. \eqref{eq:fe_scalar}, $\varphi=\textbf{\rm const}$ is not a solution of the field equations in this case and compact objects are always endowed with a scalar field. 

The particular normalization used in eq. \eqref{eq:coupling_f3} is related to the fact that the leading order expansion with respect to $\varphi$ is $f(\varphi)\sim \varphi$. Thus, it is more straightforward to separate the contribution from the dimensional part of the coupling $\lambda$ and the dimensionless parameter $\beta$. Often in the literature, though, a bit different normalization is considered, namely \cite{Pani:2011xm, Kleihaus:2014lba,Kleihaus:2016dui}
\begin{equation}
    f_{\rm EdGB_2}(\varphi) = \frac{1}{4}e^{2 \beta \varphi}.
    \label{eq:coupling_f4}
\end{equation}
Clearly, this is a very slight modification of the coupling function that can be absorbed by rescaling the parameter $\lambda$ in the following way  $\lambda \rightarrow (\lambda \sqrt{\beta/2})$. Due to practical reasons only, we give also some of the final theory constraints in terms of the coupling \eqref{eq:coupling_f4}.

The second type of coupling function is associated with sGB theories admitting spontaneous scalarization having $\frac{df}{d\varphi}(0)=0$ and $\frac{d^2f}{d\varphi^2}(0)\neq0$. The former condition secures that $\varphi=\textbf{\rm const}$ is a solution of the field equations. The latter offers a mechanism of destabilizing the GR-like compact object (for strong enough spacetime curvature) giving rise to a nontrivial scalar field configuration. The couplings we will focus on are
\begin{equation}\label{eq:sGB_f1}
    f_{SS_1}(\varphi) = -\frac{1}{2\beta}(1- e^{-\beta\varphi^2}),
\end{equation}
\begin{equation}\label{eq:sGB_f2}
    f_{SS_2}(\varphi) = \frac{1}{2\beta}(1- e^{-\beta\varphi^2}).
\end{equation}
Even though the two differ just by a minus sign, the solution properties are very distinct.  While neutron stars can scalarize for both $f_{SS_1}$ and $f_{SS_2}$, static black holes can develop a nontrivial scalar field only for $f_{SS_2}$. The coefficients in the coupling function are adjusted in such a way that the leading order expansion is $\pm \varphi^2$.

\section{Methodology}
\label{Methodology}
In the present paper, we combine the two most relevant methods for setting constraints on the scalar-Gauss-Bonnet gravity through neutron star observations related to the maximum neutron star mass and the orbital decay of binary compact objects. Both constraints are related to pulsar timing observations because the most massive neutron star up to date is observed as a pulsar in a binary.

\subsection{Constraints through maximum neutron star mass}
The first approach is very straightforward and refers to the observation that the branches of solution (possessing nontrivial scalar field) in sGB gravity have typically smaller maximum mass compared to GR. The reason for this is twofold. First,  the scalarized branches of solutions are terminated at some finite central energy density due to violation of the regularity condition \eqref{eq:SolutionConstrain}  that depends on the parameters $(\lambda,\beta)$ \cite{Pani:2011xm}. Thus, it can easily happen that the branch is terminated before the appearance of a turning point in the mass-central energy density dependence that can significantly lower the allowed maximum mass. Even if this does not happen for the chosen combination of $(\lambda,\beta)$ and EoS, one should take into account the following. For a given central energy density a GR neutron star has typically larger masses than its scalarized counterpart. Thus, the maximum mass of the sGB neutron stars, for a given equation of state, is smaller. 

\begin{figure}[]
	\includegraphics[width=0.45\textwidth]{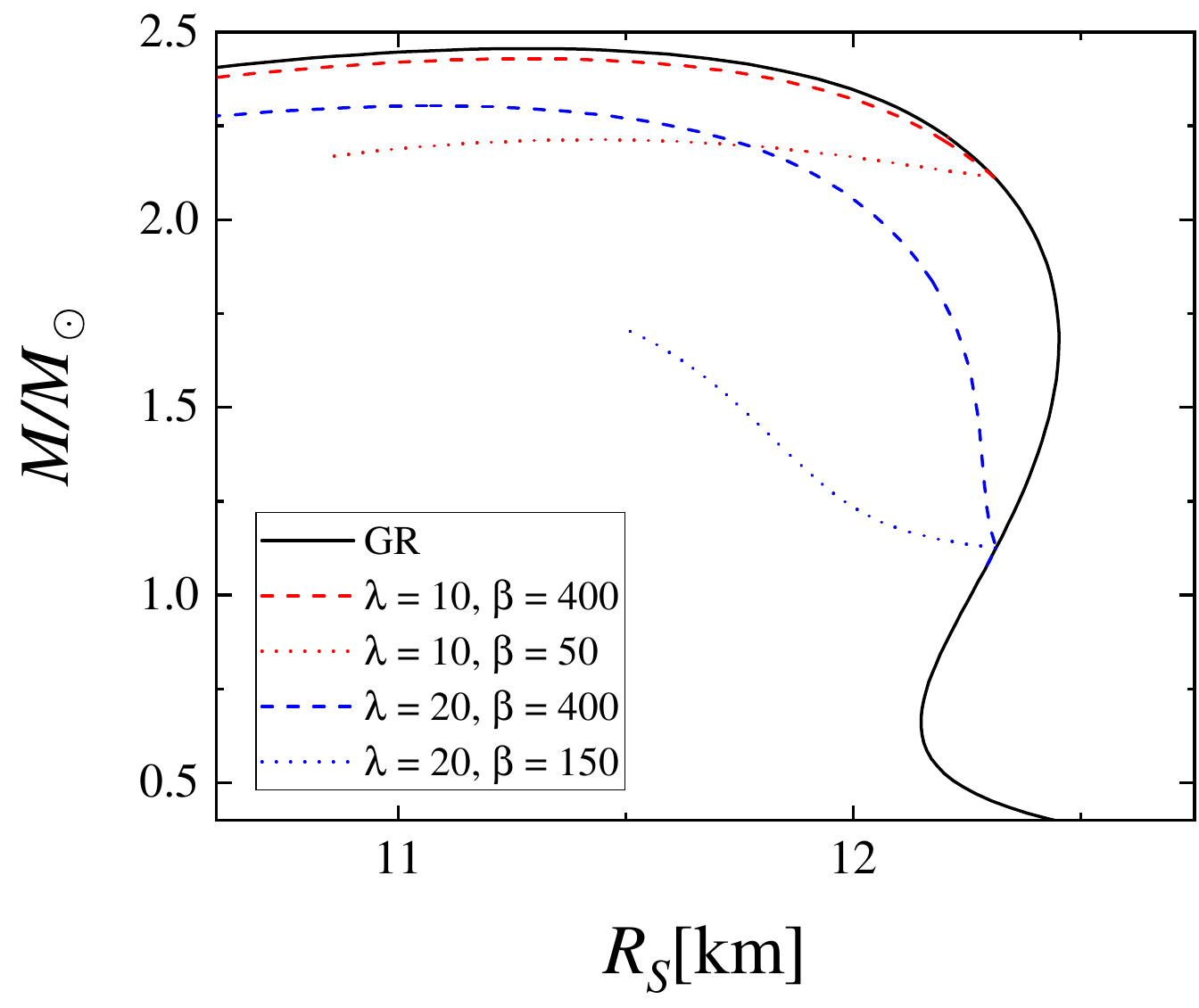}
	\includegraphics[width=0.45\textwidth]{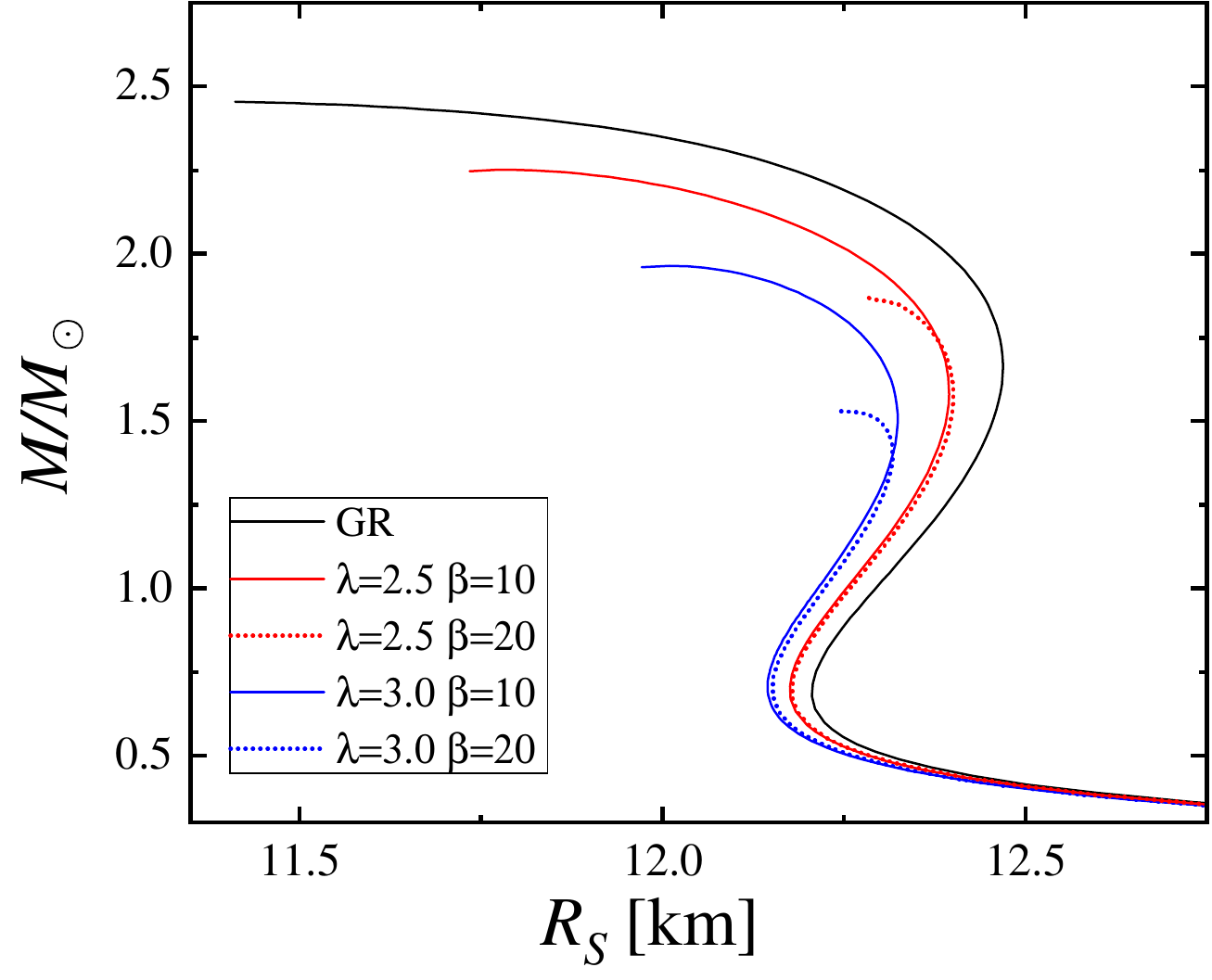}
	\caption{Mass of radius relation for sequences of neutron stars with the MPA1 EoS in sGB gravity. The theory parameters are chosen to demonstrate the different types of branch behavior and their influence on the maximum mass.(\textit{left}) Gauss-Bonnet gravity with scalarization (coupling function $f_{SS_1}$). (\textit{right}) Einstein-dilaton-Gauss-Bonnet gravity (coupling function $f_{\rm EdGB_1}$).   }
	\label{fig:M(R)}
\end{figure}

For better understanding, in Fig. \ref{fig:M(R)} we present a mass of radius relation for the two types of couplings discussed above. The left panel depicts sGB gravity with spontaneous scalarization (coupling function $f_{SS_1}$ \eqref{eq:sGB_f1}), while the right panel -- the EdGB theory (coupling function $f_{\rm EdGB_1}$ \eqref{eq:coupling_f3}). The parameters are chosen to better visualize the different types of branches -- the ones terminated at relatively small maximum mass and the ``longer'' ones reaching a turning point but still having a maximum mass below GR. As one can notice in the case of spontaneous scalarization (left panel) there is a bifurcation point below which no scalarized solutions exist and the only stable solution is the GR one. For a given EoS, this bifurcation point depends only on the Gauss-Bonnet coupling constant $\lambda$ and not on $\beta$. In the case of EdGB gravity (right panel), no bifurcation point is present and GR is not a solution of the field equations. 

To constrain sGB theory one should simply require that a neutron star solution with the observed maximum mass is allowed for a given set of theory parameters. Of course, the results will be also EOS-dependent. This approach has already been employed in \cite{Pani:2011xm} where constraints on a combination of $\lambda$ and $\beta$ were derived. However, newer observations supply us with an updated larger maximal neutron star mass that will eventually set tighter constraints on the theory. In addition, the scalar field coupling functions $f(\varphi)$ we consider are more general compared to \cite{Pani:2011xm}. 

\subsection{Constraints through orbital decay of binaries}
The second more sophisticated method for probing sGB gravity is based on the observed orbital decay of the binary pulsars. In sGB gravity the shrinking of the orbit will be sped up by the emission of scalar dipole radiation, in case of a nonzero scalar field charge. This will manifest itself as an excess in the orbital decay compared to GR. The excess can be measured through pulsar timing observation and up to now it is consistent with zero within the observational accuracy \cite{Wex:2020ald}, thus no deviation from GR can be confirmed. 

As we noted, scalar dipole radiation is emitted only for a neutron star with a nonzero scalar charge, defined as the coefficient in front of the leading order $1/r$ 
asymptotic at infinity. For a faster decaying scalar field (e.g. $1/r^2$ or an exponential decrease observed for example for a massive scalar field) the scalar charge is zero despite the presence of a strong scalar field in the vicinity of the compact object. Thus, observations of the orbital decay eventually set constraints not directly on the presence of a scalar field, but rather on the value of the neutron star scalar charge. In sGB gravity, the presence/absence of a nonzero scalar charge is controlled by the particular choices of $f(\varphi)$ and $V(\varphi)$.

Contrary to gravitational radiation which is quadrupolar, the scalar radiation has monopol, dipolar and quadrupolar components. Typically for a binary pulsar in sGB gravity, the dipolar one dominates, followed by the quadrupolar one. In our studies, we will take into account only the former one that is a good first approximation. The change in the orbital period of the binary system $P_b$ associated with the scalar dipolar radiation has the form \cite{Damour:1992we,Damour:1996ke}
\begin{equation}
    \dot{P}^{\mathrm{dipole}}_{b} = - \frac{2\pi G}{c^3} \biggl(1+\frac{e^2}{2}\biggr)(1-e^2)^{-5/2} \biggl(\frac{2\pi}{P_b}\biggr)\frac{m_p m_c}{m_p + m_c}\biggl(\frac{D_p}{m_p} - \frac{D_c}{m_c}\biggr)^2, 
    \label{eq:pbdot_dip1}
\end{equation}
where $D_p$ is the dilaton charge of the primary pulsar, $D_c$ is the dilaton charge of the companion (typically either a neutron star (NS) or a white dwarf (WD)), $m_p$ and $m_c$ are the masses of the pulsar and its companion, and $e$ is the eccentricity of the binary orbit. From the above expression, it is clear that for the dipolar emission to be strong not only the scalar charge has to be large but also the $D_p$ and $D_c$ have to be significantly different (e.g. an equal mass binary neutron star system will still have zero scalar dipole radiation). In the present paper, we will focus predominantly on cases when only the primary pulsar has scalar hair while the companion is a non-scalarized white dwarf having negligible $D_c$. Such systems give for the moment the strongest constraints on sGB gravity \cite{Danchev:2021tew}.

When interpreting observation, one should proceed in the following way. First, all kinematic effects (the relative acceleration between the binary and the Solar System barycenter along the line of sight, the Shkolovskii effect, the mass loss of the system, the tidal effects, and the possible variation of the cosmological constant on a cosmological time scale) should be subtracted from the observed total orbital decay rate ($\dot{P}_b$). This would give us the intrinsic orbital decay $\dot{P}_b^{int}$, that is the total orbital energy lost due to gravitational radiation (denoted by $\dot{P}_{\mathrm{b}}^{\mathrm{GR}}$ with a dominant quadrupolar contribution) and potentially scalar radiation (mainly the scalar dipolar one $\dot{P}^{\mathrm{dipole}}_{b}$). When the gravitational contribution is subtracted from the intrinsic orbital decay, the result is the so-called ``excess'' orbital decay $\dot{P}_b^{xs}$ 
\begin{equation}
\label{Pb_dot_xs}
    \dot{P}_{\mathrm{b}}^{\mathrm{xs}}  = \dot{P}_{\mathrm{b}}^{\mathrm{int}} - \dot{P}_{\mathrm{b}}^{\mathrm{GR}}.
\end{equation}
Thus, measuring $\dot{P}_b^{xs}$ will give us an upper limit on the scalar dipolar radiation and eventually constrain the scalar charges through eq. \eqref{eq:pbdot_dip1}.

The ultimate approach one can follow is to perform a  Bayesian analysis taking into account all observational uncertainties in the observed quantities such as $\dot{P}_b^{xs}$, the pulsar mass, etc. \cite{Danchev:2021tew} (or even perform a more sophisticated data analysis \cite{Batrakov:2023imd}). Since simpler classical methods \cite{Damour:1996ke,Freire:2012mg} have shown comparable good performance to the Bayesian analysis for the DEF model \cite{Anderson:2019eay} we were motivated to develop a procedure similar to \cite{Damour:1996ke} but adjusted for sGB gravity. 
Namely, for each EoS we build a two-parametric family of solutions in a $(\lambda, \beta)$ plane, having a constant mass equal to the mass of the observed pulsar. Practically speaking, for each combination of $\lambda$ and $\beta$ we search for a central energy density $\rho_c$ that produces a neutron star model with the desired mass. For each neutron star model in this two-parametric family of solutions, one can compute the dipole radiation through eq. \eqref{eq:pbdot_dip1} and compare it with the orbital decay excess $\dot{P}_b^{xs}$ for the given binary. Clearly, there exists a line that separates regions where $\dot{P}^{\mathrm{dipole}}_{b}>\dot{P}_b^{xs}$ and vise versa. Practically speaking, this line gives us a $\beta$ dependent constraint on the parameter $\lambda$ (or vice versa). This procedure is much more straightforward and requires much less computational effort compared to a Bayesian analysis.

\section{Results}
Before starting with the presentation of our results, let us summarize the data for the binary systems we are going to study. In Table \ref{Param_table} we list the main parameters of the binary pulsar systems we adopt in the orbital decay study. Those are the three systems studied in \cite{Danchev:2021tew} with one additional system that was found to also provide good constraints of EdGB gravity \cite{Yagi:2015oca}. We strive for consistency with \cite{Danchev:2021tew} because we are using their statistical results as a reference for testing our methodology. 

\begin{table}[!h]
\begin{center}
\footnotesize
\begin{tabular}{|c|c|c|}
 \hline
  Quantity & J0348+0432 values & J1012+5307 values \\
 \hline
 Orbital Period ($P_b$) in days   & $0.102424062722 \pm 7 \times 10^{-12}$   &$0.60467271355 \pm 3 \times 10^{-11}$\\
 Eccentricity ($e$) &   $2.6 \times 10^{-6} \pm 9 \times 10^{-7}$  & $1.2 \times 10^{-6} \pm 3 \times 10^{-7}$\\
 Intrinsic $\dot{P}_{\mathrm{b}}^{\mathrm{int}}$ in ($\mathrm{fs.s^{-1}}$) &$-274 \pm 45$ & $-2.1 \pm 8.6$\\
 NS to WD mass ratio $q \equiv m_p/m_c$    &$11.70 \pm 0.13$ & $10.44 \pm 0.11$\\
 Pulsar mass $m_{\mathrm{p}}^{\mathrm{obs}}$ in $M_{\odot}$&  $2.0065^{+0.0755}_{-0.0570}$  & $1.72^{+0.18}_{-0.17}$\\
 Observed WD mass $m_{\mathrm{c}}^{\mathrm{obs}}$ in $M_{\odot}$ & $0.1715^{+0.0045}_{-0.0030}$  & $0.165 \pm 0.015$\\
 \hline
  Quantity & J2222-0137 values & J1738+033 values \\
 \hline
 Orbital Period ($P_b$) in days   &  $2.445759995471 \pm 6 \times 10^{-12}$ & $0.3547907398724(13)$\\
 Eccentricity ($e$) &  $3.8092  \times 10^{-4} \pm 4 \times 10^{-8}$& $(3.4 \pm 1.1)\times 10^{-7}$\\
 Intrinsic $\dot{P}_{\mathrm{b}}^{\mathrm{int}}$ in ($\mathrm{fs.s^{-1}}$) & $-10 \pm 8$& $-25.9 \pm 3.2$  \\
 NS to WD mass ratio $q \equiv m_p/m_c$    &  n/a & $8.1 \pm 2$\\
 Pulsar mass $m_{\mathrm{p}}^{\mathrm{obs}}$ in $M_{\odot}$& $1.81 \pm 0.03$ & $1.46^{+ 0.06}_{-0.05}$\\
 Observed WD mass $m_{\mathrm{c}}^{\mathrm{obs}}$ in $M_{\odot}$ &$1.312 \pm 0.009$& $0.181^{+0.008}_{-0.007}$\\
 \hline
\end{tabular}
\caption{\label{Param_table} Parameters of the NS-WD pairs that were used to constrain the theory using the $\dot{P}_b$ method \cite{Antoniadis_2013, Lazaridis:2009kq, Guo:2021bqa, MataSanchez:2020pys, Ding_2020, Freire:2012mg}.  }
\end{center}
\end{table}
The heaviest known neutron star up to now is the pulsar in the binary system $J0952-0607$, discovered in 2016 \cite{Bassa:2017zpe}. It is a ``black widow'' pulsar having a sub-solar mass faint companion of minimal brightness.  Its mass is estimated at $M_{NS} = 2.35 M_{\odot}$ \cite{Romani:2022jhd}. Of course, there are uncertainties associated with this measurement and the data analysis shows that the pulsar mass should be $ > 2.09 M_{\odot}$ at $3\sigma$ confidence level. This is the value we adopt in the derivation of the maximum mass constraints on the theory. $J0952-0607$ is a relatively recently discovered pulsar, though, and up to our knowledge there is no available data for the orbital decay rate of the system in the literature. We should note that the lower limit of $2.09 M_\odot$ that we employ is almost identical to the median mass of $2.08 M_{\odot}$ of $J0740+6620$, the second massive known pulsar system \cite{Fonseca:2021wxt}.

For consistency with previous studies, the results in the rest of this section are presented in dimensionless units. The dimensionless Gauss-Bonnet coupling constant is defined as
\begin{equation}
    \lambda \rightarrow \frac{\lambda}{R_0},
\end{equation}
where $R_0 \sim 1.4766$ km is one-half of the gravitational radius of a one solar mass object. The constant $\beta$ in the coupling functions is dimensionless by definition.

\subsection{Constraints on sGB gravity admitting scalarization}

We start our study with the case of sGB gravity admitting spontaneous scalarization. This case has already been extensively studied in \cite{Danchev:2021tew} where Bayesian analysis was applied and the results for three binary pulsars and multiple equations of state are presented. In this section, we aim to demonstrate that the simpler methodology described above, and borrowed from the DEF model, is applicable for sGB gravity as well and there is a good agreement with the results from the statistical analysis. For comparison reasons, we focus on the two coupling functions eqs. \eqref{eq:sGB_f1} and \eqref{eq:sGB_f2}, that were employed also in  \cite{Danchev:2021tew}. The scalar field radial profile is very different for both cases \cite{Doneva:2017duq,Kuan:2021lol} and consequently the binary pulsar constraints differ as well \cite{Danchev:2021tew}. That is why it is important to consider both. Since we aim in this section to prove that the method can be applied and has similar accuracy to the statistical one, only one pulsar ($J0348+0432$ from Table \ref{Param_table} that provides the most stringent constraints) and two equations of state are presented, namely MPA1 \cite{Muther:1987xaa} and WFF1 \cite{Wiringa:1988tp} that are representative examples allowing a maximal mass in GR above $2.09 M_{\odot}$

\begin{figure}[]
	\includegraphics[width=0.45\textwidth]{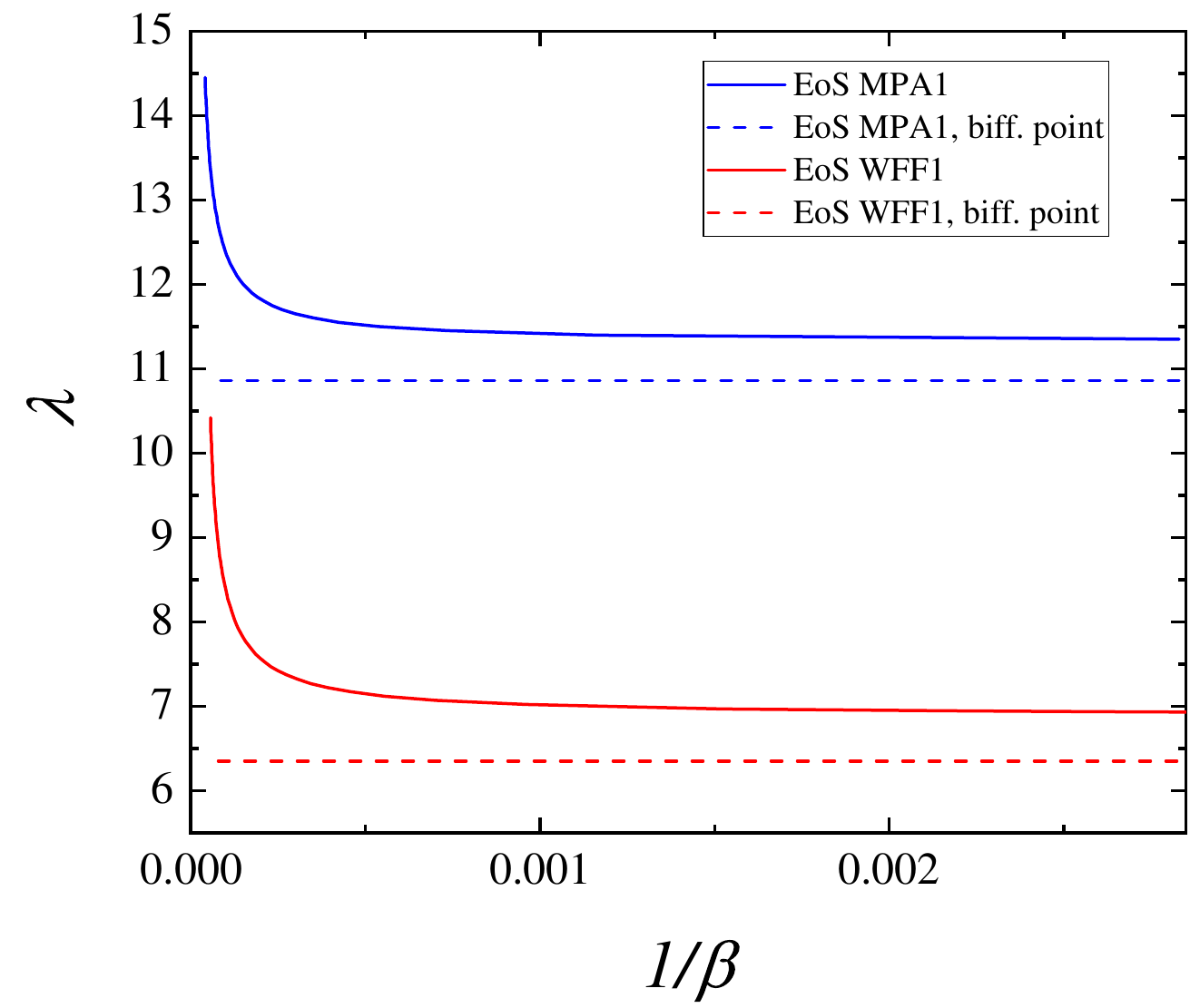}
	\includegraphics[width=0.45\textwidth]{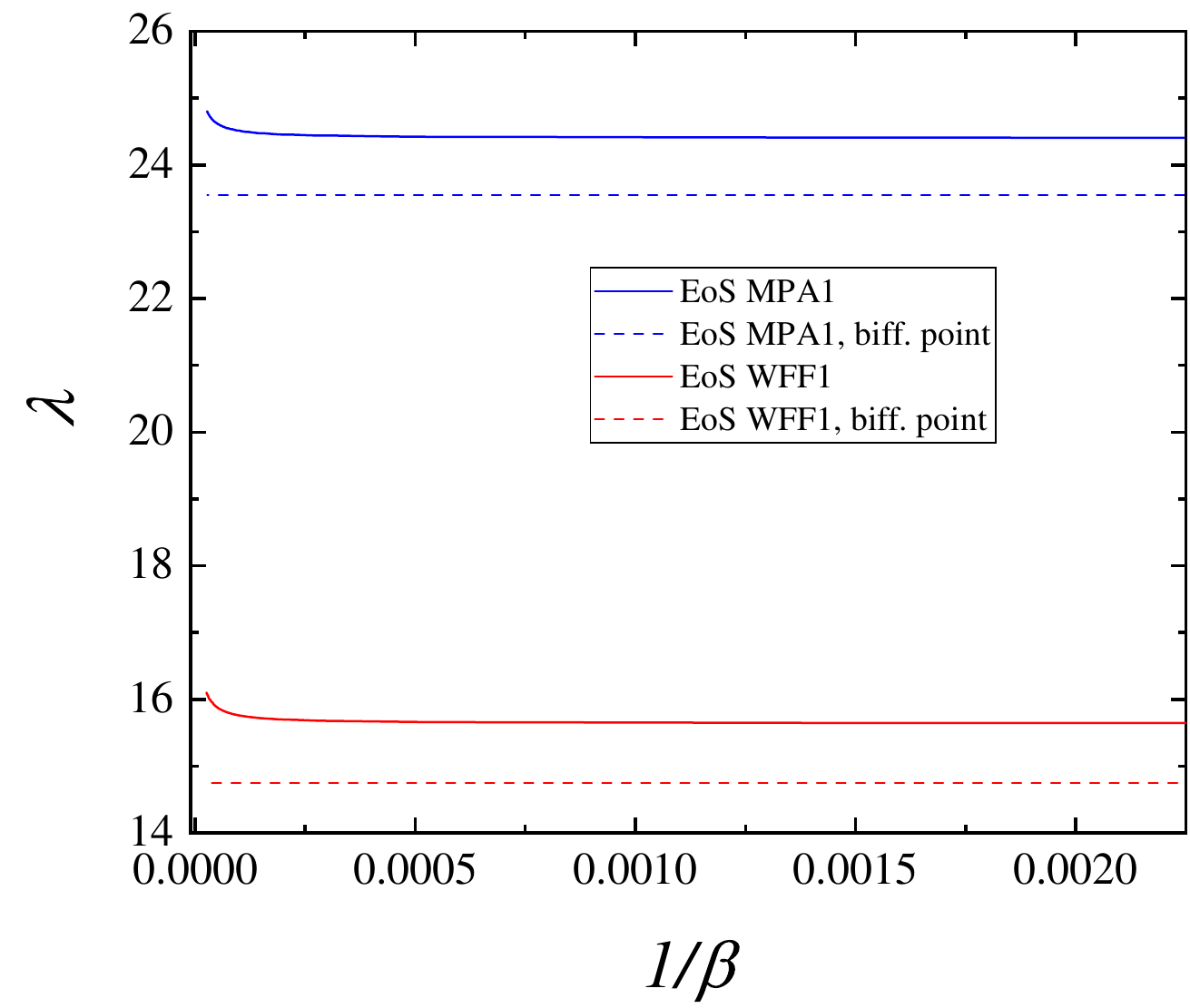}
	\caption{The constraints on sGB gravity admitting scalarization, obtained from the orbital decay rate of $J0348+0432$ for two equations of state (depicted with solid lines). The dashed line denotes the threshold value of $\lambda$ below which scalarized solutions exist only for neutron star masses above the pulsar one and thus no constraints can be imposed. This limit is related to the bifurcation points in the left panel of Fig. \ref{fig:M(R)} and is independent of $\beta$.  \textit{(left)} Coupling function (\ref{eq:sGB_f1}). \textit{(right)} Coupling function (\ref{eq:sGB_f2}). }
	\label{fig:sGB}
\end{figure}

The results for the binary pulsar constraints on the parameters $(\lambda, \beta)$ are given in Fig. \ref{fig:sGB}. In both panels, the dashed line marks the critical value of $\lambda$ at which the bifurcation point of the scalarized branch from the GR one is exactly at a mass equal to the mass of the pulsar. This line is independent of $\beta$. For $\lambda$ below this line, no scalarized solutions with the desired mass exist and therefore, this region is not constrained from observations. The continuous line is the critical curve at which the scalar dipole radiation is equal to the excess in the orbital decay. The area between the dashed and the continuous lines is that part of the parameter space for which scalar dipole radiation is present, but it is lower than the excess. Summing up, the values of parameters situated below the solid line are allowed from observations. Moreover, we have explicitly checked, that for all considered values of $\lambda$ and $\beta$ the constraints coming from the maximum observed neutron star mass are weaker than the one in Fig. \ref{fig:sGB}. 

Let us describe in more detail how the binary data is used and how we got those curves. For the bifurcation line we used the median mass of the pulsar. This line is not used to set an actual contain on the theory (at least for the coupling functions considered in the present subsection) but instead just indicates where scalar dipole radiation exists at all.  For the continuous line the procedure is the following. The scalar dipole radiation given by eq. \eqref{eq:pbdot_dip1} depends on the scalar charge as well as the parameters of the system -- the orbital period, the eccentricity and the masses of the pulsar and its companion.  All of the parameters of the system, though, are known with some uncertainty (Table \ref{Param_table}). 
The uncertainty of the orbital period is negligible and can be ignored. The eccentricity is very small by itself and gives little contribution to eq.\eqref{eq:pbdot_dip1}. Therefore, using the median value from Table \ref{Param_table} and ignoring its uncertainty will leave our results practically unchanged. The mass of the pulsar, though, has a major impact on the dipole radiation, and varying it within the uncertainty interval leaves a clear effect on the theory constraints. Since typically a larger mass would give stronger constraints (while keeping the rest of the parameters fixed) the most orthodox approach is to employ in our studies the lower limit for the pulsar mass, that is $m_p = 1.9495 M_{\odot}$. As far as the mass of the companion is concerned, what is important is to keep the mass ration fixed because this is the parameter that is observed with a very high accuracy. Thus, we will work with $m_c = 0.1668 M_{\odot}$. The orbital decay excess is calculated as the intrinsic orbital decay from Table \ref{Param_table} minus the orbital decay predicted by GR from \cite{Antoniadis_2013}, that is $\dot{P}_{\mathrm{b}}^{\mathrm{GR}} = -258^{+8}_{-11} \mathrm{fs.s^{-1}}$. Through that procedure, one obtains that the maximal possible excess should be $\dot{P}_{\mathrm{b}}^{\mathrm{xs}} = 72 \mathrm{fs.s^{-1}}$. 

When compared with the results from the Bayesian analysis in \cite{Danchev:2021tew} it is clear that the constraints derived by the two methods are almost identical, regarded, the statistical method provides more thorough information and probability for the parameters. This shows that the methodology developed for the DEF model is applicable to sGB gravity as well, and at the same time, it is significantly less demanding from a computational point of view.

\subsection{Constraints on Einstein-dilaton-Gauss-Bonnet gravity}
The second class of sGB gravity we will consider is the so-called EdGB theory with an exponential coupling function given by eq. \eqref{eq:coupling_f3}. Often in the literature, only the leading order expansion with respect to $\varphi$ is considered, making $f(\varphi)$ a linear function of $\varphi$. This corresponds to the shift symmetric Gauss-Bonnet theory. It was proven, though, that in this case, the scalar charge of neutron stars is identically zero. Hence, no constraints from binary pulsars can be imposed \cite{Yagi:2015oca}. This observation is not true in the general case of an exponential coupling function such as eq. \eqref{eq:coupling_f3}. Indeed, in this case, the scalar charge will be still relatively small but the accuracy of the pulsar timing observations is constantly improving and it is thus interesting to check whether constraints based on the binary orbital decay are already comparable to the most up-to-date limits coming from binary merger observations \cite{Yang:2023aak}. In addition, the constraints coming from the updated maximum neutron star mass observations \cite{Romani:2022jhd} are independent of the scalar charge, and thus they apply to the shift symmetric flavor of the theory as well.

\begin{figure}[t]
    \includegraphics[width=0.5\textwidth]{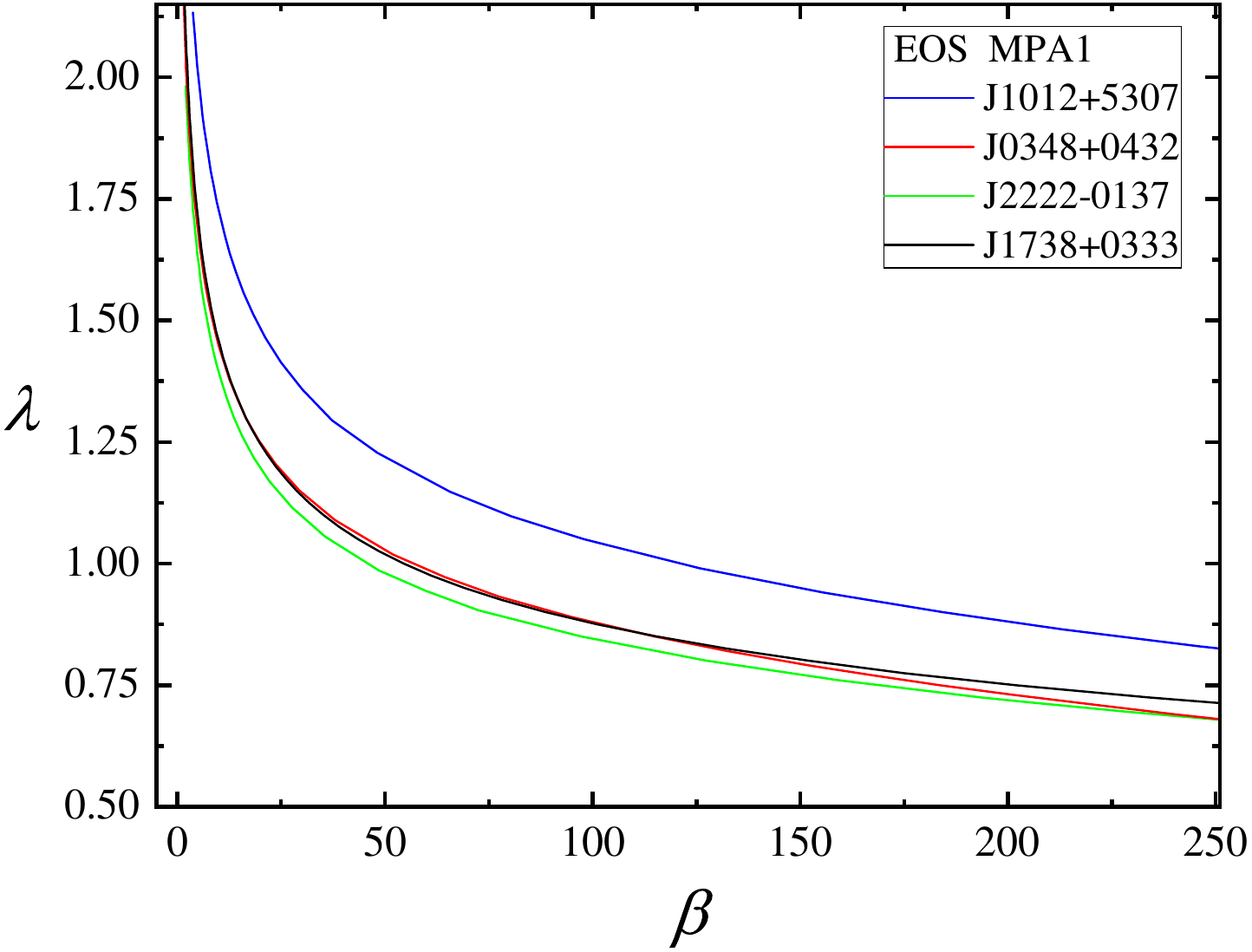}
	\caption{ Constraints on the parameters $\lambda$ and $\beta$ for four NS-WD pairs, based solely on dipole radiation due to the scalar field. The EdGB theory with coupling \eqref{eq:coupling_f3} is considered.  }
	\label{fig:EdGB}
\end{figure}

For calculating the constraints related to the orbital decay we adopted the binaries presented in Table \ref{Param_table}. We used the piecewise polytropic approximation \cite{Read:2008iy} of four different equations of state which allow for maximal masses larger than $2.09 M_{\odot}$ in GR (the lower $3\sigma$ confidence limit of the mass of $J0952-0607$), namely -- MPA1 \cite{Muther:1987xaa}, APR3 \cite{Akmal:1998cf}, APR4 \cite{Akmal:1998cf}, and WFF1 \cite{Wiringa:1988tp}.  Constraints on the parameters $(\lambda,\beta)$, using the coupling function (\ref{eq:coupling_f3}), for all three pulsars, and fixed equation of state are presented in Fig. \ref{fig:EdGB}. As expected, as $\beta \rightarrow 0$ the constraints get weaker because we get closer to the shift symmetric theory with $f(\varphi) \sim \varphi$ where the scalar charge is always zero for neutron stars. The pulsar J1012+5307 gives the least stringent constraints on the $(\lambda,\beta)$ space, while the rest of the systems ($J0348+0432$, $J2222-0137$ and $J1738+033$) lead to similar results. This is different from the couplings \eqref{eq:sGB_f1} and \eqref{eq:sGB_f2}, where the heaviest pulsar gives the most stringent constraints \cite{Danchev:2021tew}.  Instead, in EdGB gravity what plays a role is a combination of the pulsar mass and the accuracy of the $\dot{P}_b^{xs}$ measurement.

In what follows we will predominantly focus on $J0348+0432$ and $J2222-0137$ as representative examples. Let us point out that for the latter system, the white dwarf has a significantly higher mass compared to the rest. Remember that in this flavor of sGB gravity, any compact object will be endowed with a scalar field. As the compactness decreases, though, the source term for the scalar field, that is the Gauss-Bonnet invariant, decreases very rapidly. Thus, the scalar charge even for such a high-mass white dwarf (having a typical radius of at least an order to magnitude larger than a neutron star) can be safely neglected in comparison to the pulsar. 

\begin{figure}[h]
    \includegraphics[width=0.45\textwidth]{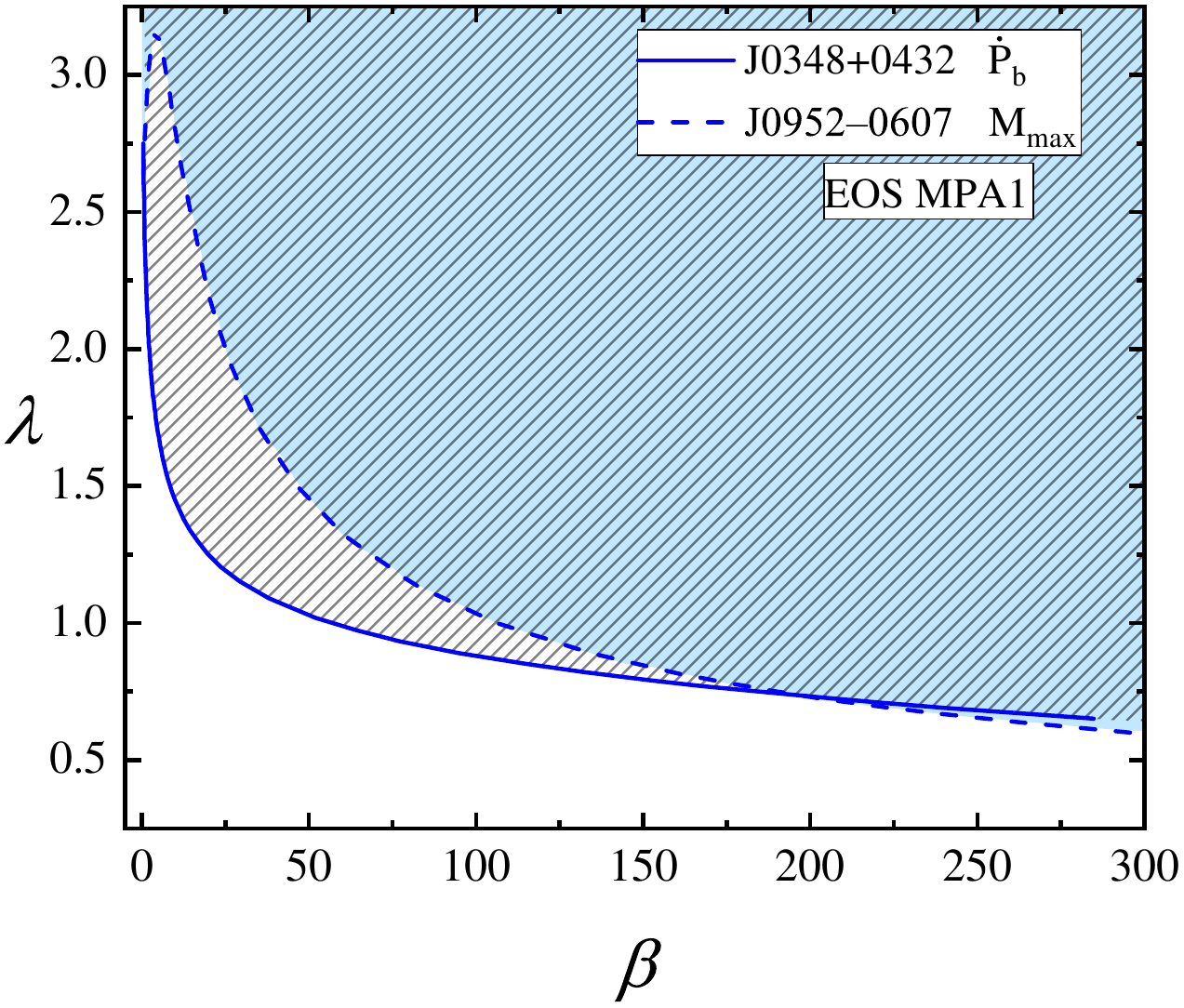}
    \includegraphics[width=0.45\textwidth]{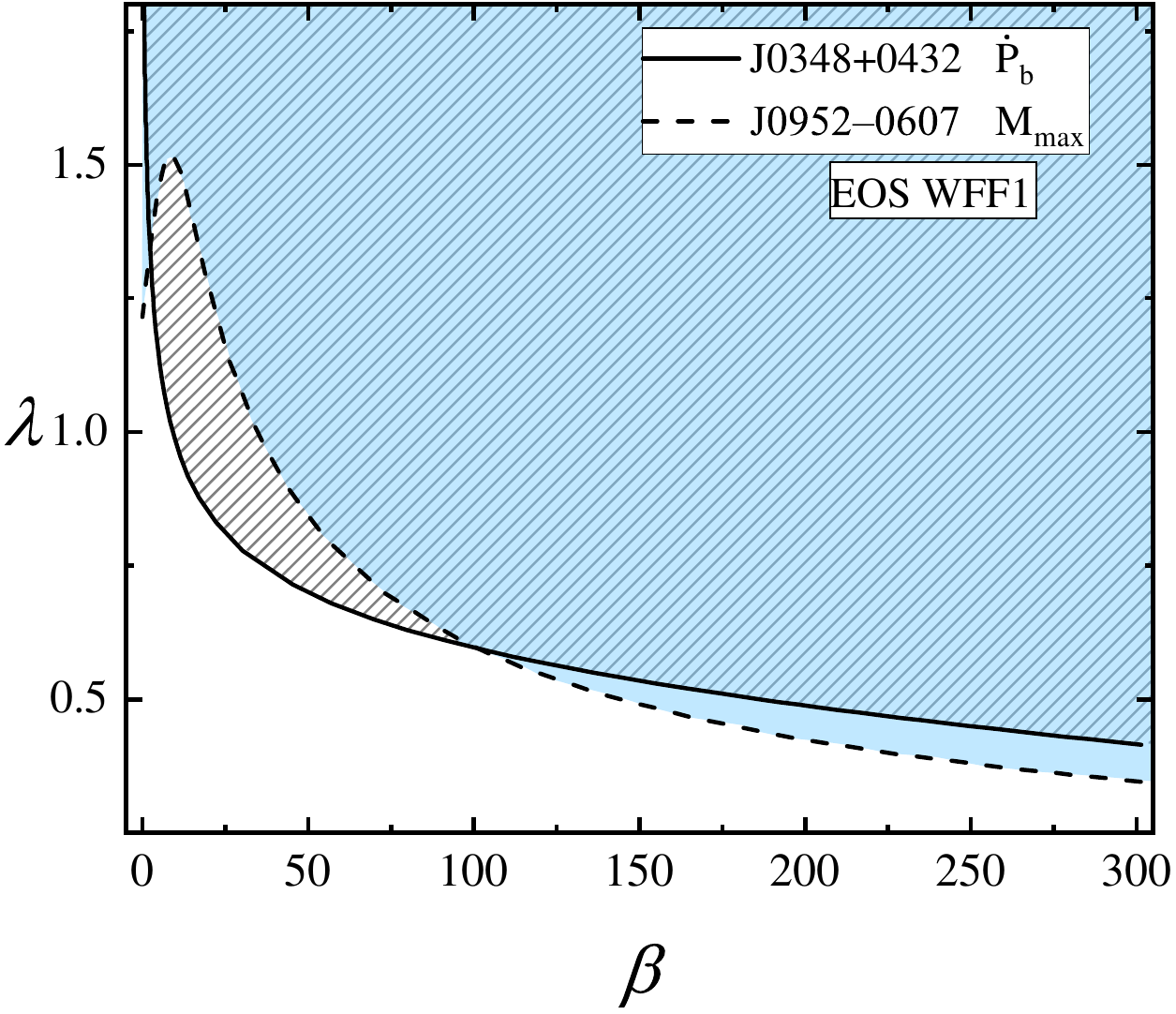}
	\caption{ Constraints on the $\lambda - \beta$ parameter space based on maximal mass and orbital decay due to the emission of gravitational radiation for two equations of state. The EdGB theory with coupling \eqref{eq:coupling_f3} is considered.  }
	\label{fig:EdGB_shade}
\end{figure}

In Fig. \ref{fig:EdGB_shade} we proceed by comparing the constraints on the $(\lambda,\beta)$ parameter space obtained by the orbital decay and the maximal mass methods for two representative EoS, namely MPA1 and WFF1, and the binary $J0348+0432$. The dashed line corresponds to combinations of $(\lambda,\beta)$ for which the maximal mass of the resulting sequence of beyond-GR neutron stars is exactly $2.09 M_{\odot}$. The solid line corresponds to the parameters for which the scalar dipole radiation for a $J0348+0432$-like system is exactly the measured $\dot{P}_{b}^{xs}$ for this binary. The shaded blue area (the region above the dashed line) indicates the set of parameters for which the maximal mass $M=2.09M_{\odot}$ cannot be reached, while the shaded grey area (the region above the solid line) -- that part of the parameter space in which the scalar dipole radiation of the system is larger than the $\dot{P}_{b}^{xs}$ for the binary.  In the white area are the parameters allowed by both methods. From the figures, it is clear that for small and high values of $\beta$ the requirement for maximal neutron mass above $2.09 M_{\odot}$ dominates, that is natural since for $\beta \rightarrow 0$ the scalar charge of the neutron stars is vanishingly small. For intermediate values of $\beta$, on the other hand, the $\dot{P}_{b}^{xs}$ measurement provides the more relevant constraints.

\begin{figure}[h]
    \includegraphics[width=0.45\textwidth]{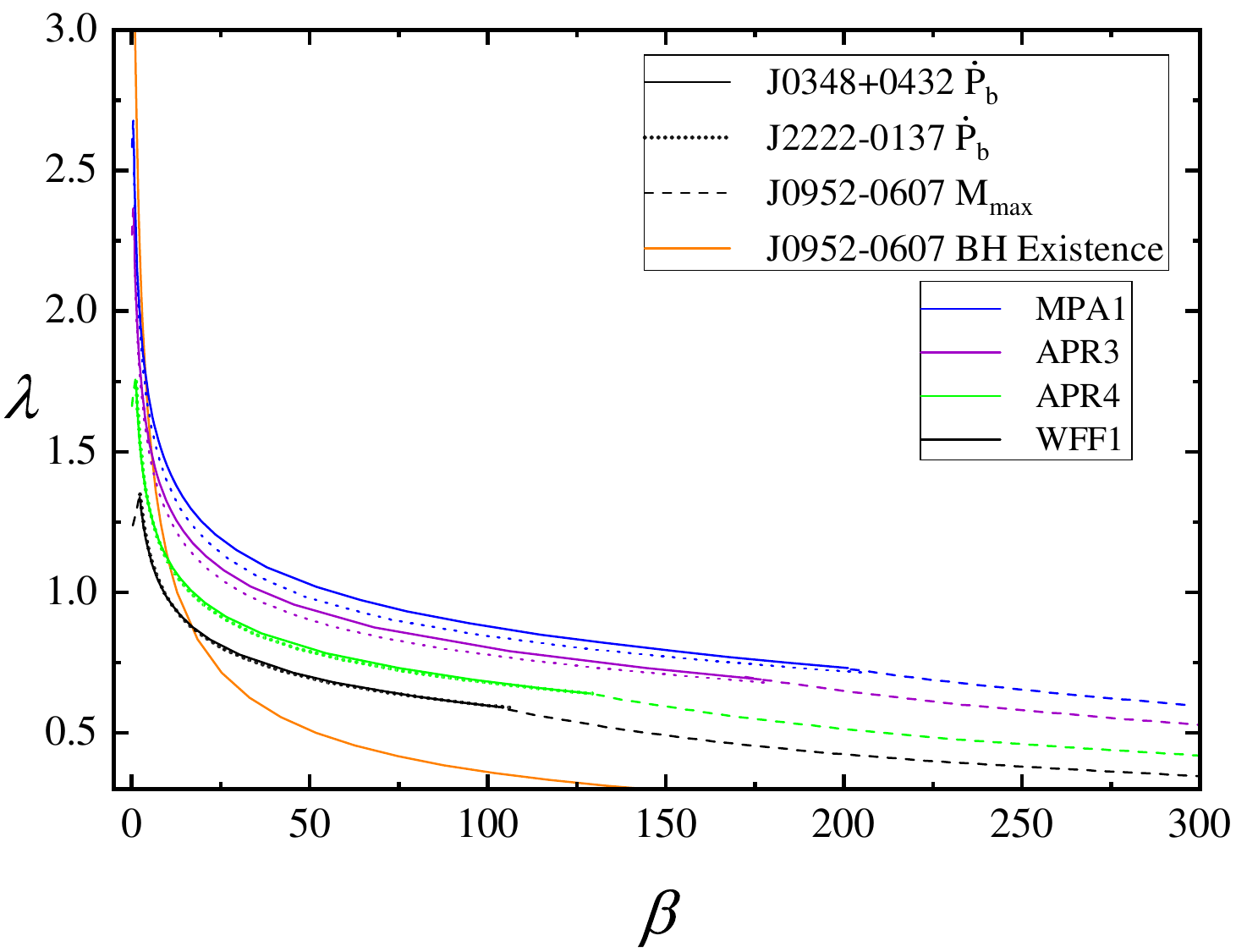}
    \includegraphics[width=0.45\textwidth]{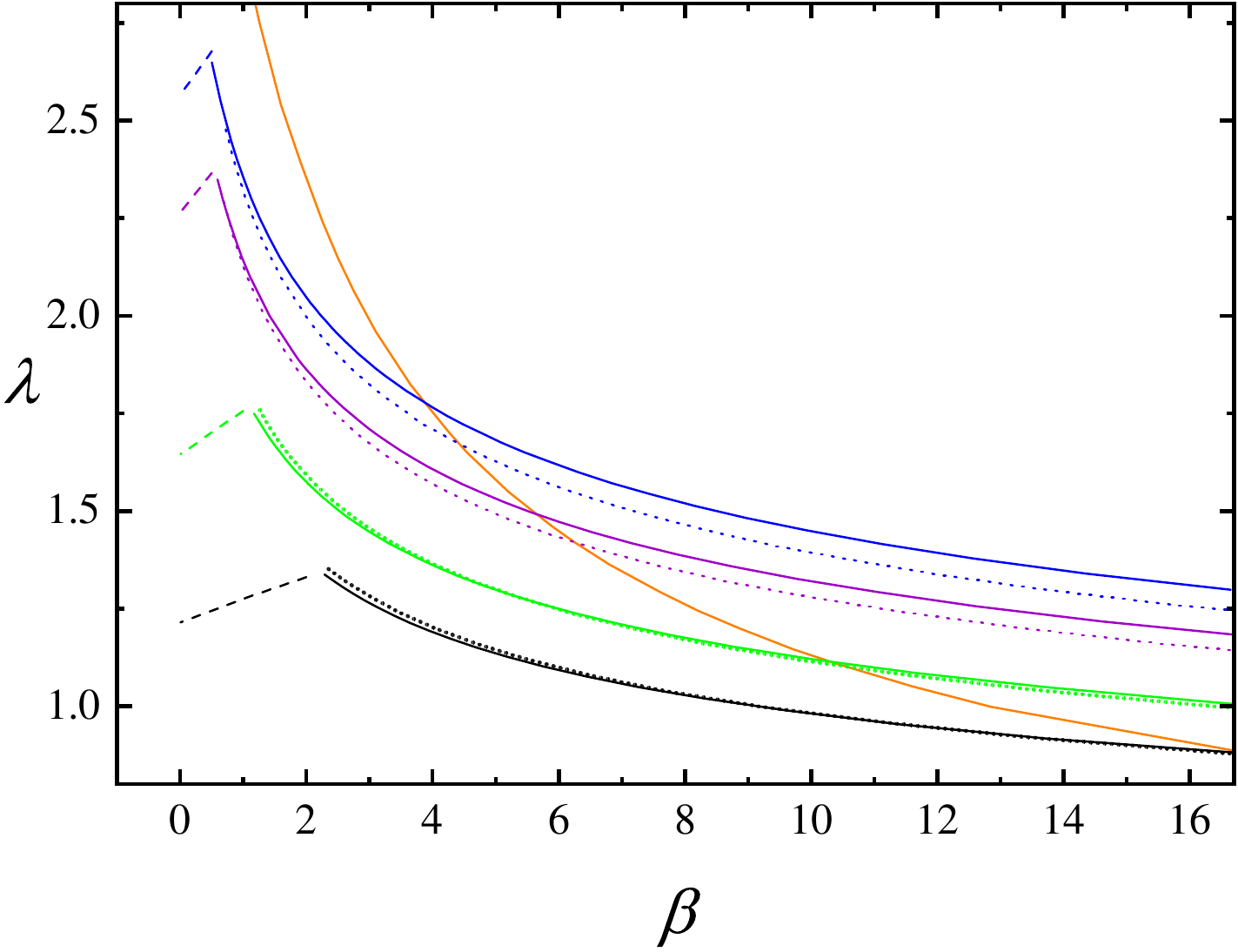}
	\caption{ Constraints on the parameters $\lambda$ and $\beta$, based on two constraining methods, using different types of observational data and for four different EoS. Results from the orbital decay for both $J0348+0432$ and $J2222-0137$ are presented for all four equations of state. The EdGB theory with coupling \eqref{eq:coupling_f3} is considered. In addition, the orange line marks the constraints coming from the existence of minimum mass black hole. \textit{(left)} Wide $\beta$ range plot. \textit{(right)} Zoom in of the constraints for small  values of $\beta$.}
	\label{fig:EdGB_Pbdot_Mmax}
\end{figure}

In Fig. \ref{fig:EdGB_Pbdot_Mmax} we study and compare the constraints on the parameters space for the four EoS, MPA1, APR3, APR4 and WFF1. Results from the orbital decay for both $J0348+0432$ and $J2222-0137$ are presented for all four equations of state.  The curves that separate the allowed from the forbidden part of the parameter space are obtained by combining the two methods: The part of the curve related to the orbital decay is marked with a continuous line and the contribution from the maximal mass method -- with a dashed line. The allowed parameters for a given EoS and binary are below the given curve. 
More or less for all values of $\beta$  the uncertainty in the EoS leads to the fact that the maximal allowed values for $\lambda$ double from the EoS with the lowest lambda to the one with the highest.

An interesting fact is that in EdGB gravity there is a minimal mass allowed for the black holes, and it depends on the parameters of the theory. This happens due to a violation of the regularity conditions at the black hole horizon that is a relation similar to eq. \eqref{eq:SolutionConstrain} for neutron stars. This by itself provides additional constraints on the parameter space since the theory parameters should be such to allow the existence of the minimum observed black hole mass. Currently, the gravitational wave observations show that the minimum black hole mass should be at least roughly $M_{\mathrm{bh}} = 5 M_{\odot}$ measured for the event \text{GW190924\textunderscore021846} \cite{LIGOScientific:2020ibl}. This constraint is plotted as an orange line in the figure, marking the combinations of $(\lambda,\beta)$ for which the minimal allowed mass for the sequence is exactly $5 M_{\odot}$. The allowed parameters are below this curve. It is clear that such a constraint strongly restricts the parameter space only for large values for $\beta$ while for small  $\beta$ the neutron star constraints prevail.

\begin{figure}[h]
    \includegraphics[width=0.45\textwidth]{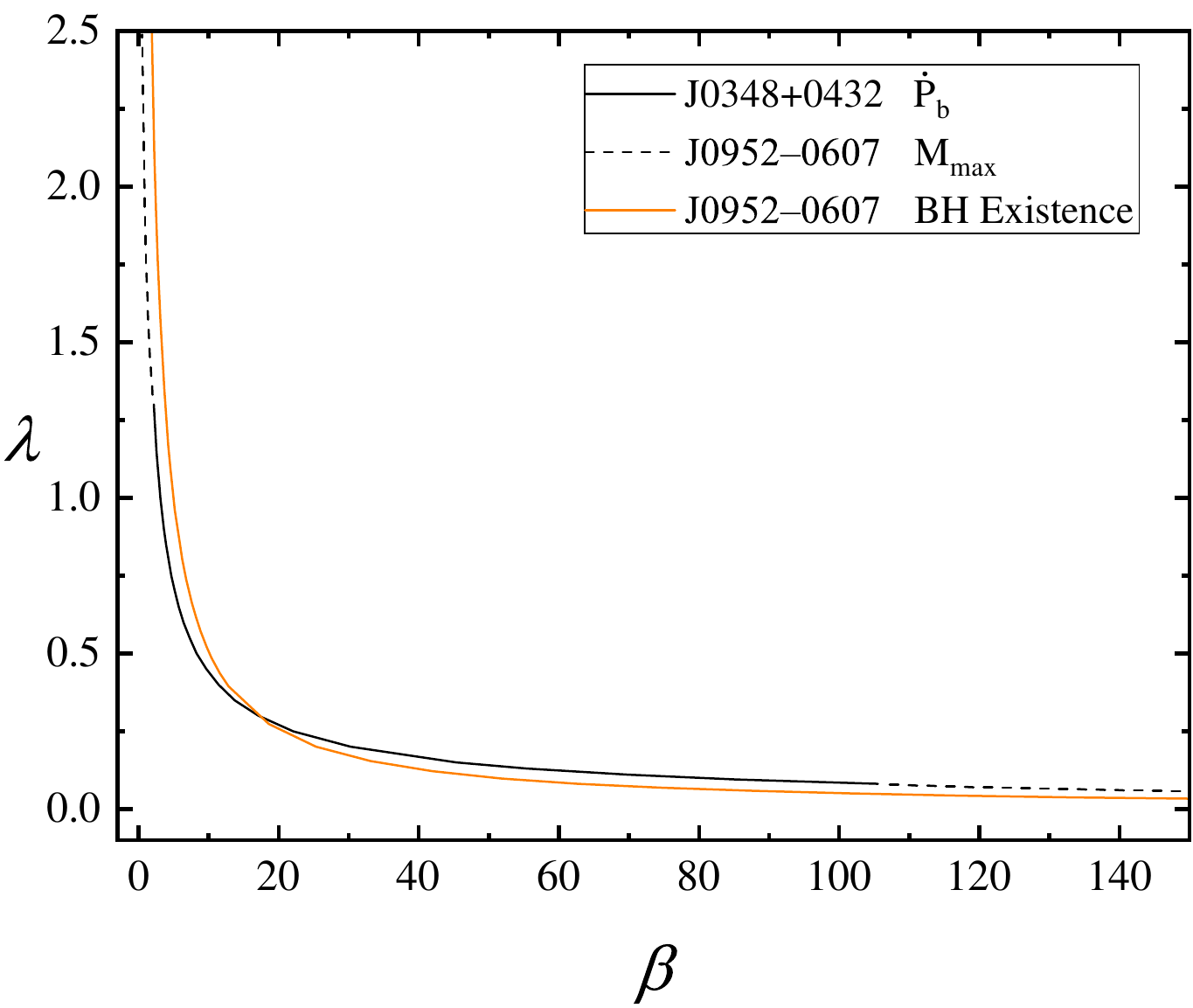}
    \includegraphics[width=0.45\textwidth]{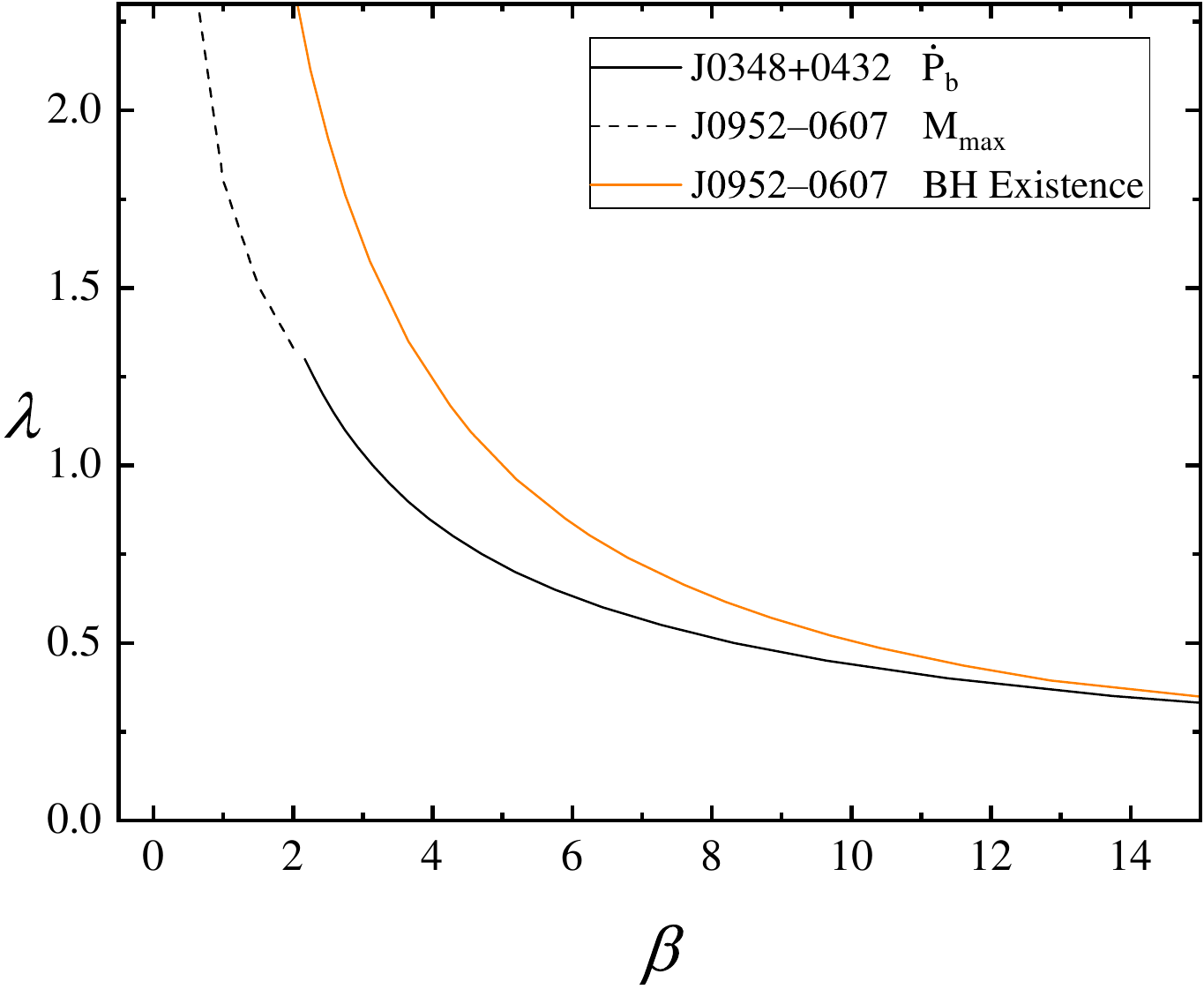}
	\caption{ Constraints on the parameter space $\lambda$ - $\beta$, similar to Fig. \ref{fig:EdGB_Pbdot_Mmax}, but for EdGB theory with coupling \eqref{eq:coupling_f4}. \textit{(left)} Wide $\beta$ range plot. \textit{(right)} Zoom in of the constraints for small  values of $\beta$.}
	\label{fig:EdGB_2pulsars}
\end{figure}

Let us now consider the coupling function (\ref{eq:coupling_f4}). As we commented, it is equivalent to the coupling \eqref{eq:coupling_f3} through a redefinition of the parameter $\lambda$. Since it is widely used in the literature, though, it will be useful to show a plot with the corresponding constraints. In Fig.  \ref{fig:EdGB_2pulsars} we present the combined constraints from the orbital decay and the maximal mass methods for $J0348+0432$ and EoS WFF1 as well as the black hole constrain for the coupling (\ref{eq:coupling_f4}). As it could be expected, the values for $\lambda$ are just rescaled with respect to the previous figures. It is interesting to point out that due to the nonlinear connection between the coupling constants, for small $\beta$ the maximal allowed $\lambda$ increases with the decrease of $\beta$, contrary to the coupling (\ref{eq:coupling_f3}).

In the end, we should see how our results compare with other available constraints in the literature. In \cite{Pani:2011xm} the authors derive constraints on the parameters of EdGB gravity with exponential coupling by using the maximal mass of neutron stars. At that time $M \geq 1.93 M_{\odot}$ so clearly, our constraints are stronger. In addition, we used the full form of the exponential coupling function, while in \cite{Pani:2011xm} only the limit of small scalar fields and thus weak coupling was considered.
Recently, the maximal mass constraints on the sGB gravity were updated in the case of shift-symmetric Gauss-Bonnet gravity \cite{Saffer:2021gak} with a newer value for the maximal mass $M = 2.01 M_{\odot}$. The most stringent constraint the authors report in that paper is for EoS MPA1 which we use as well in the present work. They find (translated to our notations and dimensionless units) $\lambda < 2.53$ that is similar to our results in the $\beta \rightarrow 0$ limit. This minor inconsistency can easily be explained by the fact that in \cite{Saffer:2021gak}, the authors use perturbative methods to solve the field equations, while we solve the field equations numerically with no simplifications. We should also note that for the selected set of equations of state, the constraints from the equation of state MPA1 are the least stringent ones.

 In \cite{Yagi:2015oca} the authors derive constraints on EdGB gravity with exponential coupling from scalar dipole radiation from neutron star-white dwarf binaries. They employ  Tolman VII and polytropic $n=0$ models. In addition, the parameter in the coupling function exponent, denoted by $\gamma$ in their notations, is set to $1$. When translated to our notations and dimensionless units, $\gamma=1$ corresponds to our $\beta \sim 0.14$, and their final theory constraints for $\lambda$  are between $\sim 2.5$ and $\sim 5.1$. Those constraints are weaker than our results and the reason is that for such small values of $\beta$ the dominating constraints come from the maximal masses rather than scalar dipole radiation.

Similar constraints comes from a recent work \cite{Lyu:2022gdr} where the authors study the constraints on EdGB gravity by analyzing the gravitational wave signal from black hole -- neutron stars binaries. They work in the small coupling approximation which results in shift-symmetric Gauss-Bonnet gravity ($f(\varphi) = \varphi$). It is important to mention that constraints coming from the orbital decay are not possible in that case due to a zero scalar charge and, therefore, zero scalar dipole radiation. The maximum neutron star mass constraints are valid, though, in this limit. 
In order to compare the results we need to rewrite our action in the form they use or vice versa. This requires a redefinition of the scalar field and a redefinition of the constants.  Their combined bound $\sqrt{\alpha} \lesssim 1.18$ km translates to $\lambda \lesssim 4.44$ km in our notation or $\lambda \lesssim 3.01$ in the dimensionless units presented on the figures. It is clear that depending on the equation of state the maximal values for the coupling constant we get from the maximal mass constraint is between less than half of what they get up to values very similar to theirs. Concerning the constraints we get from the minimal black hole mass, the maximal values for the coupling constant is in correlation with their result.

\section{Conclusion}
In the present paper, we aimed to explore methods for constraining sGB gravity through observational data from binary pulsar systems. As a first step, we adjusted a simplified non-statistical method for imposing constraints via observations of the orbital shrinking of binary pulsars. It is based on the idea of calculating the theoretical prediction of scalar dipole radiation for a given beyond-GR neutron star model and comparing it to the excess in the orbital decay. It was carefully demonstrated that it gives comparable good results to the previous more sophisticated Bayesian analysis in sGB gravity admitting spontaneous scalarization. The second major constraint related to binary neutron stars is the observed maximal neutron star mass. Namely, the theory parameters should be adjusted in such a way so that such a high-mass neutron star exists in our theory. Additionally, we included the constraints from the lowest observed black hole mass.

Focusing on Einstein-dilaton-Gauss-Bonnet theory, we should point out that contrary to the known results in the literature, we did not explicitly imposed from the beginning small field or small coupling approximations but instead employed the full exponential form of the coupling. This allows us to properly study the two-dimensional parameter space made of the dimensional Gauss-Bonnet coupling constant $\lambda$ and the  parameter $\beta$ in the coupling function exponent.  On the other hand, this means that no single value constraint on $\lambda$ can be set but instead, it is dependent on $\beta$. In addition, the equation of state plays a major role and the resulting constraints strongly depend on it. Limiting ourselves to some of the modern, widely accepted equations of state, though, we show that the end results vary by less than a factor of two.

A general observation is that EdGB gravity is best constrained by the maximum mass method for very large and very small values of the parameter $\beta$, while binary pulsar orbital decay provides the best limits for intermediate $\beta$. We should point out that for the considered coupling $\beta \rightarrow 0$ tends to the shift symmetric sGB theory with a linear coupling with respect of the scalar field. In that case, the scalar charge and thus the scalar dipole radiation are identically zero. Hence, the  orbital decay approach can not constrain the theory and it is natural that the maximum mass observations give the strongest limits there.

Very importantly, the derived constraints are either comparable or better than the ones coming from binary mergers \cite{Lyu:2022gdr}, with an improvement up to a factor of two depending on the equation of state. This is a very intriguing results because neutron stars are often overlook as a probe of EdGB gravity.

Future observations will improve the precision of the parameters employed in the paper. Potentially, a higher maximum mass neutron star might be observed or the accuracy in the mass determination of the currently most massive pulsar can be improved. As far as the orbital decay is concerned, continuous observations of already known pulsars improve more and more the current bounds on the orbital decay excess. Therefore, we can expect that in the next decade, the bounds derived in the present paper can be significantly improved. On the other hand, the binary merger observations are improving as well, and detecting a longer inspiral phase preceding the binary merger, due to increased sensitivity or a favorable high signal-to-noise ratio event, will improve the theory constraints derived in \cite{Lyu:2022gdr}. 

\section{Acknowledgment}
We would like to thank Kent Yagi, Norbert Wex, and Paulo Freire for reading the manuscript and useful suggestions.
This study is in part financed by the European Union-NextGenerationEU, through the National Recovery and Resilience Plan of the Republic of Bulgaria, project No. BG-RRP-2.004-0008-C01. DD  acknowledges financial support via an Emmy Noether Research Group funded by the German Research Foundation (DFG) under grant no. DO 1771/1-1.

\bibliography{bibliography}

\end{document}